\documentclass[prl,twocolumn]{revtex4}

\usepackage{amsmath}
\usepackage{graphicx}
\usepackage{times}

\begin{document}

\title{A Line of Critical Points in $2+1$ Dimensions: Quantum Critical Loop Gases
and Non-Abelian Gauge Theory}

\author{Michael Freedman$^1$, Chetan Nayak$^{1,2}$, Kirill Shtengel$^{1,3}$}
\affiliation{$^1$Microsoft Research, One Microsoft Way,
Redmond, WA 98052\\
$^2$ Department of Physics and Astronomy, University of California,
Los Angeles, CA 90095-1547\\
$^3$ Department of Physics, California Institute of Technology, Pasadena, CA 91125}

\date{\today }

\begin{abstract}
We (1) construct a one-parameter family of lattice models of interacting spins;
(2) obtain their exact ground states; (3) derive a statistical-mechanical
analogy which relates their ground states to $O(n)$ loop gases;
(4) show that the models are critical for $d\leq \sqrt{2}$, where $d$
parametrizes the models;
(5) note that for the special values $d=2\cos(\pi/(k+2))$,
they are related to doubled level-$k$ SU(2) Chern-Simons theory;
(6) conjecture that they are in the universality
class of a non-relativistic SU(2) gauge theory; and 
(7) show that its one-loop $\beta$-function vanishes for
all values of the coupling constant, implying that it is also
on a critical line.
\end{abstract}

\maketitle

\paragraph{Family of Microscopic Lattice Models.}
Consider the following models of $s=1/2$ spins
on the links of a honeycomb lattice, inspired by
a model of Kitaev \cite{Kitaev97}:
\begin{multline}
\label{eqn:d-isotopy-Ham}
H_{\rm d-iso} =  {\sum_v} \biggl(1+{\prod_{i\in{\cal N}(v)}}{\sigma^z_i}\biggr)\\
+ {\sum_p} \biggl( \frac{1}{d^2}{\left({F^0_p}\right)^\dagger}{F^0_p} + 
{F^0_p}{\left({F^0_p}\right)^\dagger}
- \frac{1}{d}{F^0_p} - \frac{1}{d}{\left({F^0_p}\right)^\dagger}\\
+ {\left({F^1_p}\right)^\dagger}{F^1_p} + {F^1_p}{\left({F^1_p}\right)^\dagger}
- {F^1_p} - {\left({F^1_p}\right)^\dagger}\\
+ {\left({F^2_p}\right)^\dagger}{F^2_p} + {F^2_p}{\left({F^2_p}\right)^\dagger}
-  {F^2_p} - {\left({F^2_p}\right)^\dagger}\\
+{\left({F^3_p}\right)^\dagger}{F^3_p} + {F^3_p}{\left({F^3_p}\right)^\dagger}
- {F^3_p} - {\left({F^3_p}\right)^\dagger}
\biggr)
\end{multline}
where ${\cal N}(v)$ is the set of 3 links neighboring vertex $v$, and 
\begin{eqnarray}
{F^0_p} &=& {\sigma^-_1}{\sigma^-_2}{\sigma^-_3}{\sigma^-_4}{\sigma^-_5}{\sigma^-_6}\cr
{F^1_p} &=& {\sigma^+_1}{\sigma^-_2}{\sigma^-_3}{\sigma^-_4}{\sigma^-_5}{\sigma^-_6}
+ \mbox{cyclic perm.}\cr
{F^2_p} &=& {\sigma^+_1}{\sigma^+_2}{\sigma^-_3}{\sigma^-_4}{\sigma^-_5}{\sigma^-_6}
+ \mbox{cyclic perm.}\cr
{F^3_p} &=& {\sigma^+_1}{\sigma^+_2}{\sigma^+_3}{\sigma^-_4}{\sigma^-_5}{\sigma^-_6}
+ \mbox{cyclic perm.}
\end{eqnarray}
if $1,2,\ldots,6$ label the six edges of plaquette $p$.
This Hamiltonian is a sum of projection operators with
positive coefficients and, therefore, is positive-definite.
The eigenvalues of the first term in (\ref{eqn:d-isotopy-Ham})
are $2, 0$, corresponding to whether
there is an even or odd number of ${\sigma^z}=-1$ spins
neighboring this vertex. The zero eigenvalue corresponds
to the latter case. When eigenvalue zero is obtained at every
vertex, the ${\sigma^z}=1$ links form connected loops.
Hence, the zero-energy subspace of the first term is spanned
by all configurations of multi-loops. Loops on the honeycomb lattice cannot cross.

The term on the second line is a projection operator which annihilates
a state $|\Psi\rangle$ if the amplitude for all of the spins on a given plaquette
to be up is a factor of $d$ times the amplitude for them to all be down,
i.e. if the amplitude for a configuration with a small loop encircling a single plaquette
is a factor of $d$ times the amplitude for an otherwise identical
configuration without the small loop.
The other three lines of the Hamiltonian vanish on a state $|\Psi\rangle$
if it accords the same value to a configuration if a loop
is deformed to enclose an additional plaquette. It is useful to think
of these multi-loops as continuous curves, in which case, the ground state
is invariant under smooth deformation of a curve and it loses
a factor of $d$ if a small, contractible curve is erased. Together,
these conditions have been dubbed `$d$-isotopy' \cite{Freedman04a}.

\paragraph{Exact Ground State Wavefunctions.}
The terms on the final four lines of (\ref{eqn:d-isotopy-Ham})
do not commute with each other, but they are compatible in
the sense that they all annihilate the ground state,
which is a superposition of all configurations $\alpha$ of
multi-loops weighted by a factor of $d$ to the number of loops $n_\alpha$ in
each configuration:
\begin{equation}
\label{eqn:ground-state}
|{\Psi_0}\rangle = {\sum_\alpha} d^{n_\alpha}\,|\alpha\rangle
\end{equation}
On an $L\times L$ torus, the ground state degeneracy is $\sim L^2$
because the Hamiltonian does not mix states $|\alpha\rangle$ with
different winding numbers. The different ground states are given
by (\ref{eqn:ground-state}) but with the sum over $\alpha$ restricted
to a single topological class. More generally, the ground state 
on any genus $g\geq 1$ surface is infinitely degenerate in the thermodynamic limit.
At $d=1$, there is an additional condition
called the Jones-Wenzl projector which also annihilates
the ground state (\ref{eqn:ground-state}) on a topologically-trivial manifold
but mixes different winding number sectors on higher-genus surfaces.
Hence, at $d=1$, there are two Hamiltonians
which have the same ground state on the sphere but one of
them, given in eq. \ref{eqn:d-isotopy-Ham}, is infinitely degenerate on
the torus while the other (with the Jones-Wenzl projector added) has finite
degeneracy \cite{Kitaev97}. The second leads to a topological phase
with an energy gap \cite{Kitaev97}, while 
the first, as we will see later, is gapless
and critical. At intermediate length
scales, the two systems are the same and their physics
is determined by $d$-isotopy; it is only at longer scales
(where, for instance, the topology of the manifold
becomes apparent) that, in the latter Hamiltonian,
the critical behavior crosses
over to that of the stable phase.

While the ground state can be obtained exactly, excited states
cannot because the different projection operators on
the final four lines of (\ref{eqn:d-isotopy-Ham})
do not commute with each other. We will obtain some information
about excited states using a variational ansatz but, in order to do this,
we need to learn a little more about the structure inherent
in the ground state. 

\paragraph{Mapping of the Ground-State to a Statistical-Mechanics Problem.}
Many properties of the ground-state wavefunction can
be obtained by observing that the norm of the ground state
is equal to the partition function of a classical $O(n)$ loop
model with $n=d^2$:
\begin{eqnarray}
\label{eqn:sum-over-configs}
\langle {\Psi_0}|{\Psi_0}\rangle &=& {\sum_\alpha} {d^{2{n_\alpha}}}
= Z_{\rm O(n)}(x=n)
\end{eqnarray}
For integer $n$, this model can be defined by the partition function
\begin{equation}
\label{eqn:Boltzmann}
Z_{\rm O(n)}(x) = \int {\prod_i} {d \hat{S}_i}\,\prod_{\langle i,j\rangle}
(1+x{\hat{S}_i}\cdot {\hat{S}_j})
\end{equation}
where $\hat{S}_i$ lies on the unit $(n-1)$-sphere.
The Hamiltonian $-\beta H=\sum_{\langle i,j\rangle}\ln(1+x{\hat{S}_i}\cdot {\hat{S}_j})$
has been chosen so that its high-temperature (small $x$) series expansion
takes the form
\begin{equation}
\label{eqn:O(n)-def}
Z_{\rm O(n)}(x) = {\sum_\alpha} \left(\frac{x}{n}\right)^{\ell_\alpha}\,{n^{{n_\alpha}}}
\end{equation}
where ${\ell_\alpha}$ is the total length of the loops in the
configuration $\alpha$. The expression (\ref{eqn:O(n)-def})
is well-defined for arbitrary $n$ and $x$, so we will take
it as the {\it definition} of the $O(n)$ loop model \cite{Nienhuis87}.
For $n<2$ and $x>{x_c}=n\sqrt{2+\sqrt{2-n}}$,
this model is in its low-temperature phase,
which is critical. Spin-spin correlation functions have
power-law decay \cite{Nienhuis87},
$\langle {S}(r) \,{S}(0)\rangle \sim {r^{-\eta_1}}$,
where ${\eta_k}= \frac{1}{4}{k^2}\,g - \frac{1}{g}(1-g)^2$ and
$0<g<1$ is given by $n=-2\cos(\pi g)$.
In this regime, the loops meander about the system,
as described by a family of exponents such as $\eta$.
(These exponents are obtained for any $\infty>x>{x_c}$, and our
results apply to the corresponding family of wavefunctions within this
universality class. Precisely the same exponents are also
obtained in the {\it critical} $q$-state Potts model  with $q=n^2$
on the square lattice \cite{Saleur87,Nienhuis87},
where loops surround clusters with equal Potts spin.
This can be exploited in constructing other lattice
models in the same universality class \cite{Freedman03}.)

However, equal-time spin-spin correlation functions such as
$\langle{\sigma^{z}_i}{\sigma^{z}_j}\rangle$ in the
original quantum model (\ref{eqn:d-isotopy-Ham}) are short-ranged in space.
Such correlation functions are related to the probability
that a loop passes through $i$ and a loop which may or may not be distinct 
passes through $j$. Such correlation functions vanish in
the $O(n)$ loop models (or the related $q$-state Potts models).
These models have a Coulomb gas representation as Gaussian height models
with background charges \cite{Nienhuis87} whose presence causes
correlators of neutral operators, such as gradients of the height (to which
the local loop density corresponds) to
vanish. Algebraic decay is possible for correlation functions of operators which
are charged in the Coulomb gas picture, but these are non-local
in terms of the spins ${\sigma^{z}_i}$ since they
measure, for instance, the probability that two
spins ${\sigma^{z}_i}$ and ${\sigma^{z}_j}$ lie
on the {\it same loop}. (At the two points $d=1,\sqrt{2}$, this can also
be seen from the fact that the ground state
on the sphere is the same -- and, therefore, has the same equal-time
correlation functions -- as that of a gapped Hamiltonian \cite{Turaev92}
which is a sum of local commuting operators and, therefore,
has correlation length zero.)
Thus, the ground state wavefunction of (\ref{eqn:d-isotopy-Ham})
has an underlying power-law long-ranged structure which is apparent in
its loop representation, but it is not manifested
in the correlation functions of local operators ${\sigma^{z}_i}$.
As we will see momentarily, this long-range structure leads
to gapless excitations for the Hamiltonian (\ref{eqn:d-isotopy-Ham})
and, therefore, long-ranged correlations in time in spite of the
lack of long-ranged correlations in space. We call such a state
of matter a {\it topological critical} or 
{\it quasi-topological phase}.

\paragraph{Low-Energy Excitations.}
In spite of the short-ranged
nature of equal-time spin-spin correlation functions
and the absence of any conservation laws for the Hamiltonian
(\ref{eqn:d-isotopy-Ham}), we can construct a variational argument
that this Hamiltonian is gapless using the criticality of non-local
correlation functions. The key observation is that
the configuration space can be divided into two regions $X,Y$ -- those
configurations which have `large' loops ($X$) and those which don't ($Y$).
Let us define `large' to mean having a linear extent (the greatest distance
between two points on the loop) which
is larger than $u L$, where $L$ is the linear size of the system
and $0<u<1$ will be choosen in a moment.
If we consider a large but finite-sized system, there is a finite
probability $p(u)$ that there will be a large loop so long as
$1<d<\sqrt{2}$. {\it Since the loop model is critical,
$p$ can depend only on the shape of the system and $u$, not on its
size}; this is the key input following from the criticality of the
$O(n)$ loop model. We choose $u$ so that $p(u)=1/2$ and
consider a trial wavefunction:
\begin{equation}
|{\Psi_1}\rangle = \frac{1}{\sqrt{Z_{\rm O(n)}}}\left({\sum_{\alpha\in X}} d^{n_\alpha}\,|\alpha\rangle
- {\sum_{\alpha\in Y}} d^{n_\alpha}\,|\alpha\rangle\right)
\end{equation}
The prefactor $1/\sqrt{Z_{\rm O(n)}}$ normalizes the wavefunction.
The condition $p(u)=1/2$ guarantees that $\left\langle {\Psi_0}|{\Psi_1}\right\rangle=0$.
To compute the energy of this wavefunction, we note that
the operators $F_p^{1,2,3}$ can change the spatial extent
of a loop by at most one plaquette, i.e. by less than $2a$ where
$a$ is the lattice spacing.
Hence, the Hamiltonian $H_{\rm d-iso}$ has the special property
that $\langle\alpha'| H|\alpha\rangle=0$ for $\alpha\in X$
and $\alpha'\in Y$ unless $\alpha'\in \partial Y$
(this is necessary but not sufficient), where $\partial Y$ is
the set of loop configurations whose largest loop has linear
extent $R$ satisfying  $uL-2a<R<uL$.
If the matrix element is non-zero, it takes the value $-1$.
These considerations, along with the fact that
$H_{\rm d-iso}|{\Psi_0}\rangle=0$, tell us that
\begin{multline}
\label{eqn:variational-energy}
\left\langle {\Psi_1}| \:H_{\rm d-iso}\: |{\Psi_1}\right\rangle =\\
 -\frac{4}{Z_{\rm O(n)}}{\sum_{\alpha\in X}}{\sum_{\alpha'\in Y}} 
d^{n_{\alpha'}} d^{n_\alpha}\,\left\langle\alpha'|H_{\rm d-iso} |\alpha\right\rangle\\
\leq \frac{4}{Z_{\rm O(n)}} {\sum_{\alpha'\in \partial Y}} d^{2n_\alpha'}
\:\:= \:\: 4\left(p\left(u-{2a}/{L}\right)-p(u)\right)\\
\approx 4p'(u)\,\frac{2a}{L}
\end{multline}

\begin{figure}[tbh]
\includegraphics[width=3.45in]{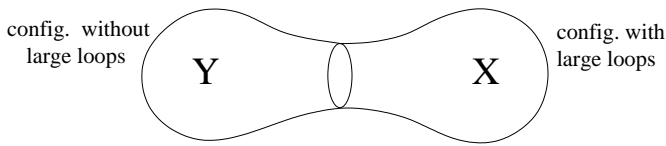}
\caption{The configuration space of the system has the
shape of a dumbbell with two regions $X$, $Y$ which
are not directly connected by the Hamiltonian. Consequently,
a trial wavefunction which has value $-\Psi_0$ on $Y$
and $\Psi_0$ on $X$ has vanishing energy in the thermodynamic
limit.}
\label{fig:dumbbell}
\end{figure}

Hence, the energy gap vanishes in the $L\rightarrow\infty$
limit for $1<d<\sqrt{2}$. The bound (\ref{eqn:variational-energy}) can be tightened
by noting that this eigenvalue problem is analogous to that for
a stretched string; by optimizing the trial wavefunction,
a more careful treatment shows that the gap vanishes with system size
as $1/L^2$. Details will be given in ref. \onlinecite{Freedman04b},
where we will also present a more general theorem which
applies to Hilbert spaces which decompose
into two subspaces $X,Y$ and Hamiltonians which,
like $H_{\rm d-iso}$, do not directly connect states in
these two subspaces, but connect them only
through a long series of repeated applications of it.
At some intermediate step in this passage from $X$ to $Y$, the system
must pass through a bottleneck; the configuration space
has the `dumbbell' form depicted in figure \ref{fig:dumbbell}.
In our case, the bottleneck
corresponds to those configurations in which the longest loop has
length $u L$. From this theorem and the analogy drawn in figure \ref{fig:dumbbell},
we see that the eigenvalue problem for our Hamiltonian is
analogous to the eigenvalue problem for the Laplacian
operator on a manifold shaped as in figure \ref{fig:dumbbell}.
The square lattice quantum dimer model at its RK point \cite{Rokhsar88}
is another example of such a system, with $X$ and $Y$ corresponding, respectively,
to configurations with large ($X$) and large negative ($Y$) total areas
(or integrated heights, in their associated height models)
enclosed by the curves in their transition graphs.

From the size-dependence of the gap, we deduce that
the low-energy excitations of $H_{\rm d-iso}$ have dispersion $\omega~\propto~k^2$.
However, these gapless modes are not Goldstone
bosons, since there is no continuous symmetry of (\ref{eqn:d-isotopy-Ham}).
They are critical modes which are unstable to the addition of
terms to our Hamiltonian which allow surgeries which cut and rejoin curves,
such as the Jones-Wenzl projectors \cite{Freedman04a}.
Such terms would directly connect the left and right sides of
figure \ref{fig:dumbbell} so that the configuration space of the system
would have the shape of a ball instead of a dumbbell and, hence,
would be gapped.

\paragraph{Conjectured Effective Field Theory.}
We do not have a direct derivation of the
low-energy effective field theory for this line
of critical points, but we can
motivate a conjecture with the following reasoning.
Quantum loop gases correspond naturally to
Wilson loops in gauge theories.
The gauge group should be SU(2) because its fundamental
representation is pseudo-real which implies -- as in the
Rumer-Teller-Weyl theorem -- that the loops are undirected.
Quantum gases of undirected loops arise in the context of
doubled SU(2) Chern-Simons topological
field theories \cite{Freedman04a}, with which our
critical points are clearly intimately related. The effective field theory
should have extensive degeneracy on the torus corresponding to the
different distinct winding numbers of the loop gas.
Finally, the effective field theory should have quadratic dispersion $\omega\propto k^2$,
rather than the linear dispersion associated with relativistic
gauge theories. The only natural choice for such a theory is
a variant of SU(2) Yang-Mills theory with the coefficient of the
usual electric field term $\text{tr\!}\left({\bf E}^2\right)$
set to zero (the absence of such a term implies the existence of infinitely-many
degenerate ground states on the torus, corresponding to different
uniform electric fields) so that the leading electric field term has
two derivatives \cite{U(1)-case}:
\begin{multline}
\label{eqn:non-rel-non-Abel-action}
S[{A^a_0},{A_i^a},{E_i^a}] =
\frac{1}{g^2}\int {d^2}x\, d\tau \biggl( {E_i^a}{\partial_\tau}{A_i^a} + 
{A^a_0} {D_i}{E_i^a}\\
-{ \frac{1}{2}\left({D_j}{E_i^a}\right)^2} + \frac{1}{2}{B^a}{B^a}
+ {\lambda_1} {\left({E_i^a}{E_i^a}\right)^2}
+  {\lambda_2} {\left({E_i^a}{E_j^a}\right)^2}
\biggr)
\end{multline}
where ${D_j}{E_i^a}~=~{\partial_j}{E_i^a} + {f^{abc}}{A_j^b}{E_i^c}$ and
the magnetic field is given by
${B^a}~=~{\partial_1}{A_2^a}-{\partial_2}{A_1^a} + {f^{abc}}{A_1^b}{A_2^c}$.
The SU(2) index $a$ takes the values $1,2,3$, while the
spatial indices $i,j=1,2$. ${f^{abc}}$ are the structure constants
of SU(2). ${A_i^a},{E_i^a}$ are an SU(2) gauge field and its
canonically conjugate electric field. The time component
of the gauge field, ${A^a_0}$, is a Lagrange multiplier
which enforces Gauss' law, ${D_i}{E_i^a}=0$.
We have written the action in first-order phase space form.
In principle, $E_i^a$ can be integrated out so that we
will have an action dependent on $A_i^a$ alone, but
this is cumbersome because $E_i^a$ and ${\partial_\tau}A_i^a$
are not linearly-related, unlike in ordinary Yang-Mills theory. 
The theory is invariant under the usual gauge transformation
${A_\mu^a} \rightarrow {A_\mu^a} + {\partial_\mu}{\alpha^a} 
+ f^{abc}{A_i^b}{\alpha^c}$, 
${E_i^a} \rightarrow {E_i^a} + f^{abc}{A_i^b}{\alpha^c}$.
Two quartic terms, with couplings $\lambda_1$, $\lambda_2$
have been included because they are marginal at tree-level.
All other higher-order terms are irrelevant or forbidden by
symmetry and have been dropped.
 
\paragraph{RG Analysis of Non-Relativistic Non-Abelian
Gauge Theory.} This theory is marginal at tree-level,
as a result of its $\omega\propto k^2$ dispersion,
which raises the concern that it might
be massive as a result of quantum fluctuations, just as $4D$ Yang-Mills is.
To address this, we compute the $\beta$-function
for this theory to one-loop using the background field method
and dimensional regularization \cite{Peskin-book}. Remarkably, it vanishes:
\begin{eqnarray}
\label{eqn:RG-eqns}
\frac{dg}{d\ell} &=& 0
\end{eqnarray}
The cancellation appears to result from the similarity between
the way in which the gauge field ${A_i^a}$ and its conjugate
momentum $E_i^a$ enter the first four terms of the action
(\ref{eqn:non-rel-non-Abel-action}), which gives us hope that
it survives to all orders. (For details, see ref. \onlinecite{Freedman04b}.)
For the same reason, the relative
scaling of space and time, which can flow in principle, is unrenormalized
at the one-loop level, so that the dynamical exponent remains $z=2$.
Since the theory (\ref{eqn:non-rel-non-Abel-action})
is on a critical line for $\lambda_{1,2}=0$,
it is a viable candidate theory for the universality class
of (\ref{eqn:d-isotopy-Ham}).
However, the one-loop RG equations
for $\lambda_{1,2}$ generically run away:
\begin{eqnarray}
\label{eqn:lambda-RG-eqns}
\frac{d\lambda_1}{d\ell} = \frac{g^2}{16}\left(24{\lambda_1} + 20{\lambda_2} -
22{\lambda_1^2}-28{\lambda_1}{\lambda_2}-7{\lambda_2^2}\right)\cr
\frac{d\lambda_2}{d\ell} = -\frac{g^2}{16}\left(8{\lambda_1} + 12{\lambda_2}
+4{\lambda_1^2}+16{\lambda_1}{\lambda_2}+14{\lambda_2^2}\right)
\end{eqnarray}
so that (\ref{eqn:non-rel-non-Abel-action}) is not critical for small $g$,
where a one-loop calculation can be trusted.
This suggests the following picture, which echoes
our earlier analysis of (\ref{eqn:d-isotopy-Ham}).
The theory is gapped for small $g$ as a result of the runaway
flow of $\lambda_{1,2}$; this corresponds to the regime $d\geq\sqrt{2}$
where (\ref{eqn:d-isotopy-Ham}) is presumably gapped. On the other
hand, we expect that for $g$ sufficiently large, $\lambda_{1,2}$ will be irrelevant
(this needs to be checked by a higher-loop calculation of the
RG equations for $\lambda_{1,2}$),
so that (\ref{eqn:non-rel-non-Abel-action}) is critical, corresponding to
the critical regime $d\leq\sqrt{2}$.
A $\text{tr\!}\left({\bf E}^2\right)\equiv{E_i^a}{E_i^a}$ term is
relevant at $g=0$, so its coefficient must be tuned
to zero. It is an important open question how relevant
this term is at the large $g$ which we expect to
correspond to $d\leq\sqrt{2}$.

\paragraph{Correlation Functions.}
At present, we do not know how to directly compute
correlation functions in the critical gauge theory
(\ref{eqn:non-rel-non-Abel-action}). However, if the correspondence
with the $d$-isotopy critical theories is correct, we can
deduce some non-trivial equal-time correlation functions
involving Wilson loop operators $W[\gamma]=
{\rm tr}\!\left({\cal P}\exp\left(i{\oint_\gamma} A\right)\right)$.
For contractible loops ${\gamma_i}$,
we expect $\left\langle W[{\gamma_1}]\,W[{\gamma_2}]\ldots W[{\gamma_n}]\right\rangle
= {d^n}$. At $g=0$, this is obtained
with $d=2$ by direct calculation.
This is somewhat surprising since
one might have naively anticipated power-law correlation
functions for fields governed by the
gapless action (\ref{eqn:non-rel-non-Abel-action}).
However, a peculiar feature of this non-relativistic
action is that many equal-time correlation functions are
short-ranged in space. For instance, dropping cubic and quartic terms,
$
\left\langle {B^a}({\bf x},0) {B^b}(0,0)\right\rangle~\sim~{g^2}{\nabla^2}\delta^{(2)}({\bf x}) \:\delta^{ab}
$.
On the other-hand, we do expect power-laws to show up in
non-local correlation functions such as:
\begin{equation}
 \left\langle {\rm tr}\left( {E_i}(x)\:
{\cal P}{e^{i{\int_0^x} A}}
\:{E_j}(0) \:{\cal P}{e^{i{\int_x^0} A}}
\right)\right\rangle \sim \frac{1}{|x|^{\eta_2}}\:\delta_{ij}
\end{equation}
where $\eta_2$ is an $O(n)$ loop model exponent defined
after (\ref{eqn:O(n)-def}).

\paragraph{Discussion.}
As we showed in the first part of this paper, the family of
Hamiltonians (\ref{eqn:d-isotopy-Ham}) is a fixed line
of critical states which are most naturally described in
the language of fluctuating unoriented loops. The Hamiltonian
is defined by topological relations which the loops must obey;
these relations can be viewed as obtained by relaxing one
of the relations (Jones-Wenzl) related to doubled
$SU(2)_k$ Chern-Simons theory, thereby leaving only $d$-isotopy.
In the second half of the paper, we found {\it another}
fixed line of critical states,
the non-relativistic SU(2) gauge theory
(\ref{eqn:non-rel-non-Abel-action}),
whose Wilson loop representation is also in terms
of  fluctuating unoriented loops.
These are the two main results of this paper. It
seems improbable that two similar remarkable
occurrences could be unrelated, so we conjecture
that the gauge theory (\ref{eqn:non-rel-non-Abel-action})
controls the infrared behavior of the universality class
containing the microscopic model (\ref{eqn:d-isotopy-Ham}).

There are a number of obvious generalizations and open
questions which we hope to address later, such as
the instabilities of these critical theories, especially
at the magic values $d=2\cos(\pi/(k+2))$;  the addition of
Chern-Simons terms to the action;
correlation functions of (\ref{eqn:non-rel-non-Abel-action})
and the precise relationship between $g$ and $d$; other gauge groups;
complex $d$ and $\theta$-terms in the action;
the effects of instantons in (\ref{eqn:non-rel-non-Abel-action});
and the implications of our results for topological quantum computation
\cite{Kitaev97,Freedman01}.

\paragraph{Acknowledgements}
We would like to thank Z. Bern, J.~M. Cornwall, P. Fendley, M.P.A. Fisher,
E. Fradkin, J. Kondev, and S. Sondhi for discussions.
We appreciate the hospitality of the Aspen Center for Physics where
part of this work was completed. We have been supported by
the ARO under Grant No. W911NF-04-1-0236.
C. N. has also been supported by the NSF under
Grant Nos. DMR-9983544 and DMR-0411800
and the A.P. Sloan Foundation.


\end{document}